# Access Control Management for Computer-Aided Diagnosis Systems using Blockchain

Mayra Samaniego[†]
Computer Science
University of Saskatchewan
Saskatoon, Canada
mayra.samaniego@usask.ca

Sara Hosseinzadeh Kassani[†]
Computer Science
University of Saskatchewan
Saskatoon, Canada
sara.kassani@usask.ca

Cristian Espana
Computer Science
AllpaTech
Saskatoon, Canada
cristian.espana@allpatech.com

Ralph Deters
Computer Science
University of Saskatchewan
Saskatoon, Canada
deters@cs.usask.ca

*Abstract*— Computer-Aided Diagnosis (CAD) systems have emerged to support clinicians in interpreting medical images. CAD systems are traditionally combined with artificial intelligence (AI), computer vision, and data augmentation to evaluate suspicious structures in medical images. This evaluation generates vast amounts of data. Traditional CAD systems belong to a single institution and handle data access management centrally. However, the advent of CAD systems for research among multiple institutions demands distributed access management. This research proposes a blockchain-based solution to enable distributed data access management in CAD systems. This solution has been developed as a distributed application (DApp) using Ethereum in a consortium network.

## I. INTRODUCTION

Artificial Intelligence (AI) has recently gained much interest in domains such as image processing, speech recognition, natural language processing, and information retrieval [1]. In the field of healthcare, the primary goal of artificial intelligence is to develop models for image analysis. The satisfying performance of machine learning, deep learning methods, and data augmentation [2] motivate institutions and companies to automate repetitive tasks [3][4]. Different systems and architectures have been developed to support AI for image recognition [5], for instance, Computer-Aided Diagnosis (CAD) systems.

Generally, CAD systems are used for decision support in the detection and interpretation of diseases, especially cancer. CAD system' pattern recognitions involve several feedback loops, which are necessary to train the system. Thus, systems can learn and produce a reliable analysis result. These feedback loops are driven by the pre-processing, segmentation, structure region of interest, evaluation, and classification of images of the affected part of the human body. These loops require and produce vast amounts of data (images and descriptive information).

This data is used to effectively train CAD systems and help clinicians provide an accurate diagnosis, start workflows, and redirect information to multiple parties. Thus, they can optimize their time when evaluating patients and prescribing treatments.

Some research has focused on developing software solutions to process the obtained datasets, for instance, analysis and processing systems [6][7][8]. Also, some works have focused on managing access to the constrained network that generates the images (e.g.: [9]).

Data access control management in CAD systems requires more focus and research. Commonly, CAD systems rely on access management models that define hierarchies, roles, rules, and privileges [10][11][12], which are controlled by a central institution infrastructure. This approach validates access to data in a centralized fashion, trusting a single institution.

This research develops a role-based access control system designed as a distributed application (DApp [13]). It implements a consortium Ethereum network to provide a decentralized architecture to handle the emerging demand for decentralized data access control management in CAD systems (e.g.: [14][15]).

The rest of the paper is organized as follows. Section two introduces CAD systems. Section three introduces access control models. Section four presents access control management in CAD systems. Section five presents the proposed system. Section six presents preliminary evaluations. Section seven presents the conclusions.

## II. COMPUTER-AIDED DIAGNOSIS (CAD) SYSTEMS

CAD systems have been developed to assist clinicians in the interpretation of medical images [1]. The output of these systems is taken as a second opinion to improve the productivity, accuracy, and consistency of clinicians [16]. CAD systems can find abnormalities from input medical images and generate diagnosis results for analysis and assessment of medical images by using computer vision and neural network algorithms leading to the improvement of automated diagnosis [1]. In general, image processing in CAD systems involves the following steps. First, pre-processing to detect bugs, reduce noise, and harmonize the quality of images. Second, segmentation to separate structures. Third, the structure of the region of interest (ROI) to focus on and analyze specific areas individually. Fourth, evaluation and classification to evaluate the possibility of a true positive.

The methodology to be used in CAD systems depend on the part of the human body that is affected. For instance, effective pre-processing steps for pigmented skin lesion images include illumination correction, contrast enhancement, color space transformation, and the removal of non-skin noises such as hair to address data heterogeneity caused by imaging instruments during acquisitions of skin images and enhance the performance and accuracy of classification and segmentation methods. The result of applying these pre-processing steps is shown in figure 1. During this process, vast amounts of data are generated.

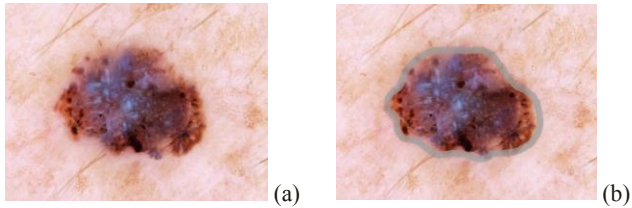

Fig. 1. Example of a processed image of a skin lesion. (a) Row image. (b) Image after application of processing steps.

CAD research has mainly focused on the development of processing algorithms and investigating the utility of the output generated by those algorithms. CAD systems have been developed to assist in different disease evaluations; for instance, Schmid-Saugeon et al. [17] present a CAD system to evaluate medical images of skin lesions. Verma et al. [18] present a CAD system to distinguish benign and malignant patterns for digital mammograms. Akram et al. [19] propose a three-stages CAD system to detect brain tumors from magnetic resonance images. However, data access control management is an area that still needs more attention. The emergence of medical projects among institutions around the world makes it necessary to introduce new forms of data access management that include the active participation of the participant institutions.

### III. ACCESS CONTROL MODELS

Access control is the process of monitoring and managing every entry to a system and its resources, including data [20].

The following are some types of access control models:

- Role-based access control (RBAC) [10]. This model defines roles to manage access permissions. This model seeks that every role has a set of specific permissions associated with it. Users must be associated with roles to get permission.
- Rule-based access control [11]. This model defines rules to manage access permissions. The entry point for evaluating those rules are external parameters, such as date time or geographical location.
- Attribute-based access control (ABAC) [12] [21]. This model also defines a set of rules to manage access permissions. However, the entry point for evaluating those rules are attributes that users define.

The characteristic that these models have in common is the need for an authority to create the roles, rules, or attributes, respectively. Traditionally, this authority has been represented as a centralized superuser, which has no restrictions. Some researchers have implemented the previously explained access models to develop access control systems (e.g., [22]).

### IV. ACCESS CONTROL MANAGEMENT IN CAD SYSTEMS

CAD systems store various types of data – for instance, compressed and uncompressed image data, and derived data structures that result from the processing of images. Some institutions employ RBAC to define roles linked to different levels of access [23]. For example, an oncologist might be allowed to have access to a patient's MRI or CT scans while a nurse might have access to the descriptive information of the patient. Roles are issued by the appropriate authority (e.g., a hospital) to determine whether or not a user (or group of users working together) should be trusted to have access to data resources.

#### A. Access Control Management in CAD Systems using Blockchain

Blockchain is a distributed ledger that can store transactions of any kind of assets and provides an immutable history of operations over data [23][24]. Blockchain has been used in different scenarios to manage access to data and resources (e.g.: [25][26][27][28]).

The combination of blockchain and RBAC can be a promising approach to address the demand for decentralized access control management in CAD systems. A blockchain-based access control system can leverage the security and privacy of information, enhancing the ethical use of CAD data. Additionally, blockchain-based access control systems can ensure a secure and reliable operation over CAD data within a collaborative project. For example, cryptographic hashes of doctors that belong to different institutions can be stored on a consortium blockchain (private blockchain network that is formed by multiple institutions) [29]. The consortium blockchain validates these hashes in a decentralized manner through a consensus mechanism to determine whether a user should be trusted to get access to data. This approach leads to actively preventing malicious intrusions and maintaining the integrity of CAD data.

The features of blockchain and DApps can bring various benefits to CAD systems such as security and protection of medical data, intellectual property rights, and ownership of tangible assets. Blockchain is a robust and reliable technology that ensures the authority of the user's credential to any data access request and thus prevents the possibility of an unauthorized user accessing private CAD data.

## V. PROPOSED SYSTEM

This research proposes a system that integrates blockchain technology for access control management in CAD systems.

### A. Use Case

It is necessary to know the specific use case in which the blockchain system will work and how the software architecture changes after its integration. Figure 2 shows the general software architecture of the CAD system. The software architecture has five macro components, which are a web interface, access control manager, view provider, artificial intelligence (AI) engine, and the evaluation factory.

The web interface implements the standard web development tools Python, JavaScript, HTML, and Flask web framework (figure 2). The access control manager implements role-based access control management (figure 3). The view provider provides views for the roles created (figures 4a shows an example of the view for an oncologist doctor and figure 4b shows an example of the view for a pharmacist). The artificial intelligence (AI) engine trains the CAD system to perform data processing. The evaluation factory processes images and gives a diagnosis as a second opinion.

### B. Design Using Blockchain

In the previously explained CAD system architecture, all software components work centrally, trusting a client-server architecture. The access control management component performs the access validation relying on a superuser. If the superuser gets corrupted, then the permissions granted cannot be trusted and have to be re-evaluated. Thus, it becomes necessary to implement an access control management process that does not depend on a single superuser role.

This work proposes an architecture in which the role-based access control management does not rely on a single superuser role. Instead, the access control management relies on a blockchain network. Figure 5 shows the software architecture of the CAD system integrating blockchain. We continue using three of the five software components, web interface, AI training engine, and evaluation factory. However, we replace the access control manager and view provider components with a blockchain network. This approach eliminates the superuser role that was the main responsible for granting/revoke access permissions. Instead of having a single superuser, we have distributed the responsibility towards a decentralized blockchain network that has to prove their work by using their computing power.

We have implemented a consortium blockchain network using Ethereum [30] to allow the participation of nodes from multiple institutions. Even though the institutions might be considered trustable, Ethereum still finds them a trust-less environment, forcing them to execute the proof of work consensus mechanism to validate their access permissions. The RBAC model has been implemented as a DApp, and the access permissions are stored in-chain

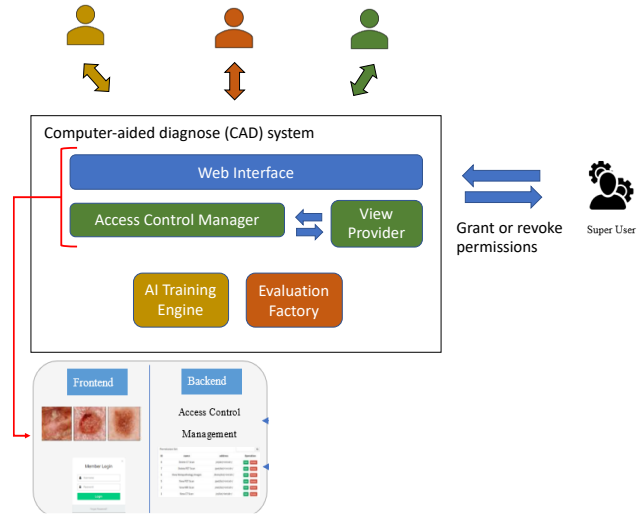

Fig. 2. General software architecture of the computer-aided diagnosis (CAD) system in which blockchain will be implemented.

Fig. 3. Example of the definition of roles and permissions for different types of medical images of the proposed system.

Fig. 4. a) Example of data delivered to an oncologist. b) Example of data delivered to a pharmacist.

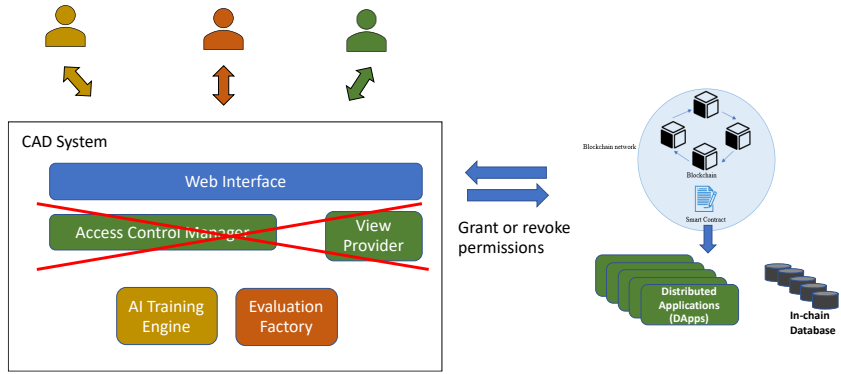

Fig. 5. General software architecture of the computer-aided diagnosis (CAD) system integrating blockchain

Figure 6 shows an example of the code of the DApp for delegating permission and restricting access. It has been programmed using solidity language.

```solidity
2   import "DelegatePermission.sol";
3
4   contract Delegated is Delegatable {
5     mapping (address => bool) public delegates;
6
7     function addDelegate (address _newDelegate) onlyOwner {
8       delegates[_newDelegate] = true;
9       accessListChanged(_newDelegate, delegateChanges.Added);
10    }
11
12    function removeDelegate (address _delegateToRemove) onlyOwner {
13      delegates[_delegateToRemove] = false;
14      accessListChanged(_delegateToRemove, delegateChanges.Removed);
15    }
```

Fig. 6. Defining smart contract rule for delegating permission.

## VI. PRELIMINARY EVALUATION

Preliminary evaluations were performed to measure the performance of the blockchain-based system that handles RBAC access control management.

Table 1 presents the characteristics of the physical nodes that were used to configure the consortium Ethereum network. We set four Ethereum nodes.

TABLE I. SPECIFICATION OF THE ETHEREUM BLOCKCHAIN NODES

| Hardware | Details |
| --- | --- |
| Operating System | Linux Debian 10.4 |
| CPU | Intel(R) Core(TM) i7-6700 CPU @ 3.40GHz |
| RAM | 16 GB |

Figure 7 shows the results of the preliminary evaluations. For the experiment, we submitted 1000 transactions organized in random groups. The CAD system redirected the operations of access validation to the blockchain network. In the graphs, the x-axis represents the blocks of transactions (100 blocks), and the y-axis represents the time it takes the blockchain system to process each block.

The preliminary results show that the RBAC blockchain system takes between 1.1 sec and 2.8 sec to evaluate each block of transactions. The fact that the blocks were sent with no delay between them might have caused that the system gets busy and spends more time when mining the blocks.

These results might vary depending on the configuration parameters of the Ethereum network, for instance, different hosting locations of nodes. Additionally, these results show that the segmentation and design of the distributed application (DApp) does not affect the overall performance of the CAD system.

## VII. CONCLUSIONS

In this research, we have implemented a blockchain-based system for RBAC management in CAD systems. The system was designed using Ethereum in a consortium network. The RBAC model was designed as a DApp. This system handles the emerging demand for distributed data access management in CAD systems. Our blockchain-based system eliminates the superuser role and distributes that responsibility towards a blockchain network.

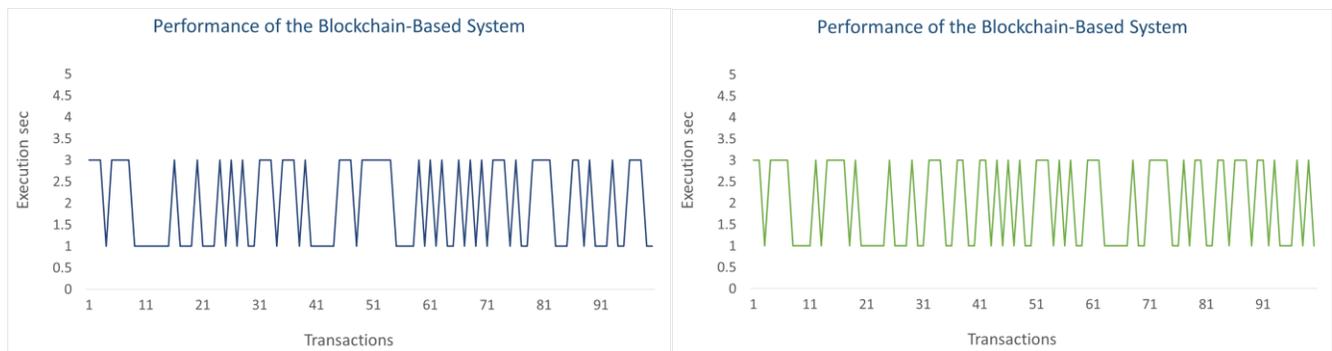

Fig. 7. Results of preliminary evaluations